\documentclass{imamat}

\usepackage{modamsthm,modnatbib}

\usepackage{latexsym,amssymb,lastpage}
\usepackage{graphicx,amsfonts}
\usepackage{times,mathptmx,bm,amsmath}
\usepackage{dcolumn}

 \newtheoremstyle{theorem}{6pt}{6pt}{\rm}{}{\sffamily}{ }{ }{}
 \theoremstyle{theorem}

  \newtheoremstyle{thm}{6pt}{6pt}{\rm}{}{\sffamily}{ }{ }{}
 \theoremstyle{thm}
\newtheorem{thm}{\sc Theorem}[section]

 \newtheoremstyle{lemma}{6pt}{6pt}{\rm}{}{\sffamily}{ }{ }{}
 \theoremstyle{lemma}
 
 \newtheoremstyle{lem}{6pt}{6pt}{\rm}{}{\sffamily}{ }{ }{}
 \theoremstyle{lem}

\newtheoremstyle{case}{6pt}{6pt}{\rm}{}{}{. }{ }{}
 \theoremstyle{case}

 \newtheoremstyle{statement}{6pt}{6pt}{\rm}{}{\sffamily}{ }{ }{}
\theoremstyle{statement}

 \newtheoremstyle{corollary}{6pt}{6pt}{\rm}{}{\sffamily}{ }{ }{}
 \theoremstyle{corollary}

  \newtheoremstyle{defi}{6pt}{6pt}{\rm}{}{\sffamily}{ }{ }{}
 \theoremstyle{defi}

  \newtheoremstyle{cor}{6pt}{6pt}{\rm}{}{\sffamily}{ }{ }{}
 \theoremstyle{cor}

\newtheoremstyle{example}{6pt}{6pt}{\rm}{}{\sffamily}{ }{ }{}
\theoremstyle{example}

\newtheoremstyle{remark}{6pt}{6pt}{\rm}{}{\sffamily}{ }{ }{}
\theoremstyle{remark}
\newtheorem{remark}{\sc Remark}[section]

\newtheoremstyle{approximation}{6pt}{6pt}{\rm}{}{\sffamily}{ }{ }{}
\theoremstyle{approximation}

\newtheoremstyle{scheme}{6pt}{6pt}{\rm}{}{\sffamily}{ }{ }{}
\theoremstyle{scheme}

\newtheoremstyle{Algorithm}{6pt}{6pt}{\rm}{}{\sffamily}{ }{ }{}
\theoremstyle{Algorithm}

 \newtheoremstyle{Remark}{6pt}{6pt}{\rm}{}{\sffamily}{ }{ }{}
 \theoremstyle{Remark}

\newtheoremstyle{Lemma}{6pt}{6pt}{\rm}{}{\sffamily}{ }{ }{}
\theoremstyle{Lemma}

\newtheoremstyle{Assumption}{6pt}{6pt}{\rm}{}{\sffamily}{ }{ }{}
\theoremstyle{Assumption}

\newtheoremstyle{Proposition}{6pt}{6pt}{\rm}{}{\sffamily}{ }{ }{}
\theoremstyle{Proposition}

\newtheoremstyle{prop}{6pt}{6pt}{\rm}{}{\sffamily}{ }{ }{}
\theoremstyle{prop}

\newtheoremstyle{rem}{6pt}{6pt}{\rm}{}{\sffamily}{ }{ }{}
 \theoremstyle{rem}

\newtheoremstyle{hypo}{6pt}{6pt}{\rm}{}{\sffamily}{ }{ }{}
 \theoremstyle{hypo}

  \newtheoremstyle{Step}{6pt}{6pt}{\rm}{}{}{ }{ }{}
 \theoremstyle{Step}

 \newtheoremstyle{lema}{6pt}{6pt}{\rm}{}{\sffamily}{ }{ }{}
 \theoremstyle{lema}

\renewcommand{\theequation}{\thesection.\arabic{equation}}

\def\citeasnoun{\cite}

\newcommand{\VV}{V}

\newcommand{\CC}{{\bf C}}
\newcommand{\EE}{{\bf E}}
\newcommand{\xx}{{\bf x}}
\newcommand{\DD}{{D}}
\newcommand{\PP}{{P}}
\newcommand{\RR}{{R}}
\newcommand{\rr}{{r}}
\newcommand{\qq}{{q}}
\newcommand{\II}{{\bf I}}
\newcommand{\KK}{{K}}
\newcommand{\HH}{{H}}
\newcommand{\UU}{{U}}
\newcommand{\bb}{{\bf b}}
\newcommand{\BB}{{B}}
\newcommand{\WW}{{W}}

\newcommand{\bx}{{\bf x}}
\newcommand{\by}{{\bf y}}
\newcommand{\bA}{{\bf A}}
\newcommand{\bF}{{\bf F}}
\newcommand{\be}{{\bf e}}
\newcommand{\bC}{{\bf C}}
\newcommand{\bQ}{{\bf Q}}
\newcommand{\bG}{{\bf G}}
\newcommand{\bs}{{\bf s}}
\newcommand{\bt}{{\bf t}}
\newcommand{ \eps}{{\epsilon}}
\newcommand{\alp}{{\alpha}}
\newcommand{\real}{{\mathbb R}} 
\newcommand{\barq}{\bar{ q}}
\newcommand{\barK}{\bar{ K}}
\newcommand{\barr}{\bar{ r}}
\newcommand{\bsig}{\mbox{\boldmath$\sigma$}}
\newcommand{\bvsig}{{\mbox{\boldmath$\varsigma$}}}

\newcommand{\Diag}{{\mbox{Diag }}}
\newcommand{ \sig}{{\sigma}}
\newcommand{\sta}{{\rm sta}}
\newcommand{ \lam}{{\lambda}}
\newcommand{\half}{\frac{1}{2}}

\newcommand{\barbsig}{\bar{\mbox{\boldmath$\sigma$}}}
\newcommand{\bxi}{{\mbox{\boldmath$\xi$}}}
\newcommand{\barbvsig}{\bar{\mbox{\boldmath$\varsigma$}}}

\newcommand{\barby}{\bar{\bf y}}
\newcommand{\barbx}{\bar{\bf x}}
\newcommand{\beps}{{\mbox{\boldmath$\epsilon$}}}

\newcommand{\vsig}{\varsigma}
\newcommand{\calP}{{\cal{P}}}
\newcommand{\calX}{{\cal{X}}}
\newcommand{\calS}{{\cal{S}}}
\newcommand{\calE}{{\cal{E}}}
\newcommand{\bff}{{\bf f}}
\newcommand{\bveps}{\mbox{\boldmath$\xi$}}
\newcommand{ \Lam}{{\Lambda}}
\def\la{\langle}
\def\ra{\rangle}

\begin{document}

\title{Canonical Duality Approach for Nonlinear Dynamical Systems}
\author{
{\sc Ning Ruan}\\[2pt]
School of Science, Information Technology and Engineering, University of Ballarat, \\[6pt]
Ballarat, VIC 3353, Australia.\\[6pt]
{\sc David Y. Gao}\\[2pt]
School of Science, Information Technology and Engineering, University of Ballarat, \\[6pt]
Ballarat, VIC 3353, Australia.\\[6pt]
{\rm [Received on ******]}}
\pagestyle{headings}
\markboth{N. Ruan and D.Y. Gao}{\rm Canonical dual approach for nonlinear dynamical systems}
\maketitle

\begin{abstract}
{This paper presents a  canonical dual approach for solving
a nonlinear population growth problem governed by the well-known logistic equation. Using the
finite difference  and least squares methods, the  nonlinear
differential equation is first formulated as a nonconvex optimization problem with unknown parameters.
We then  prove  that by the canonical duality theory, this nonconvex
problem is equivalent to a concave maximization problem over a convex feasible space, which can
be solved easily  to obtain global optimal solution to this challenging problem.
Several illustrative examples are presented.}
{logistic equation, discrete dynamical systems, global
optimization, canonical duality theory, least squares method.}
\end{abstract}

\maketitle

\section{Problems and Motivations}
The logistic equation is a model of population growth first
published by Pierre Verhulst in 1838. The
continuous version of this  model (see \citeasnoun{mbchl79}) is described by the
following first-order nonlinear differential equation:
\begin{equation}
\frac{d \bx}{dt}= \rr \bx (1- \frac{\bx}{\KK})- \CC,
\end{equation}
where $\bx( t)$ represents the numbers of individuals (population biomass) at time $t$,
the real number $\rr> 0$ is the
intrinsic growth rate of population increase,  $\KK$ is the carrying capacity, or the
maximum number of individuals that the environment can support, $\CC> 0$ represents the constant harvesting rate.

By using finite difference method, the discrete version of the logistic
equation  can be written as
\begin{eqnarray}
\bx_{t+1}&=&\bx_t+\rr  \bx_t(1-\frac{\bx_t}{\KK})-\CC_t+\EE_t,
\end{eqnarray}
where $\EE_t$ is the process error, which is critically important to
the associated dynamics equation; $\bx_t$ is the stock abundance in
year $t$.
Due to the nonlinearity, this equation has remarkable non-trivial properties
and therefore it must be handled with special methods.
It is well-known that for certain given parameters,
direct iterative methods for solving this
nonlinear equation may lead to chaotic solutions.
Such a problem was first studied on computer by the theoretical population biologist Robert May in the late 1960's.
This equation and its variants still puzzle the mathematicians.

In real-world applications,  the stock biomass $\bx$ cannot be observed directly, therefore,
an observational model  can be introduced below
\begin{eqnarray}
\hat{\II}_t=\qq \bx_t = \II_t+\eps_t,
\end{eqnarray}
where
$\hat{\II}_t$ is the estimation of  abundance,   $\II_t$ is the
abundance observed, $\qq$ is the  proportional coefficient, and  $\eps_t$
is the observation error.
Base on
this  model, the  least squares method for minimizing both the observation error $\eps_t$ and the
process error $E_t$ leads to the following optimization problem.
\begin{eqnarray}
&(\mathcal{P}_0)~~~~&\min~~\sum_{t=1}^n (\qq \bx_t-\II_t)^2+\frac{\alp}{2}\sum_{t=1}^{n-1} E_t^2\\
&& {\rm s.t.}~~~\bx_{t+1}=\bx_t+\rr
\bx_t(1-\frac{\bx_t}{\KK})-\CC_t+\EE_t,~t=1,\ldots,n-1,
\end{eqnarray}
where $\alp$ is the penalty factor. Let
\begin{eqnarray*}
\DD= \left(
\begin{array}{ccccc}
0 & 1 & \cdots & 0&0 \\
0 & 0 & \ddots & 0 &0\\
\vdots  & \vdots  & \ddots & \ddots& \vdots  \\
0 & 0 & 0 & 0& 1
\end{array}
\right)\in \real^{(n-1)\times n},
\end{eqnarray*}
and
\begin{eqnarray*}
\RR= \left(
\begin{array}{ccccc}
1 & 0& \cdots & 0&0 \\
0 & 1 & \ddots & 0 &0\\
\vdots  & \vdots  & \ddots & \ddots& \vdots  \\
0 & 0 & 0 & 1& 0
\end{array}
\right)\in \real^{(n-1)\times n},
\end{eqnarray*}
the vector form optimization problem $(\mathcal{P}_0)$ can be written as:
\begin{eqnarray*}
&(\mathcal{P}_1)~~~~&\min~~\PP_1(\bx, q, K,
r)=\|q\bx-\II\|^2+\frac{\alp}{2} \left
\|\frac{r}{K}\bx^T \bA \bx +(\DD-(1+r)\RR)\bx+ \CC  \right \|^2\\
&& {\rm s.t.}~~q>0, K>0, r>0,\\
&&\;\;\;\;\;\;\bx \in \real^n, q\in \real, K \in \real, r \in \real,
\end{eqnarray*}
where $\II \in \real^n$, $\bC\in \real^{n-1}$ are given vectors,
$\bA=\{A_{ij}^t\}=\{\delta_{ij}^t\} \in \real ^{n\times
(n-1)\times n}$, and
\begin{eqnarray*}
\delta_{ij}^t = \left\{ \begin{array}{ll}
1 & \mbox{ if } i=j=t, \\
0 & \mbox{ otherwise}.
\end{array} \right.
\end{eqnarray*}

By setting  $\by=\bx/K$,  can  equivalent population dynamics
model can be proposed in the following.
\begin{eqnarray*}
&(\mathcal{P}_2 )~~~~&\min~~\PP_2 (\by, q, K,
r)= \left \|q\by-\frac{\II}{K} \right \|^2+\frac{\alp}{2}
\left \|r \by^T \bA \by +(\DD-(1+ r)\RR)\by+ \frac{\CC}{K} \right \|^2\\
&& {\rm s.t.}~~1>\by>0,q>0, K>0, r>0,\\
&&\;\;\;\;\;\;\by \in \real^n, q\in \real, K \in \real, r \in \real.
\end{eqnarray*}
Clearly, this fourth-order polynomial optimization problem with unknown parameters
is in general nonconvex, which could
have  many  local minimizers. Due to the lacking of sufficient conditions for identifying global optimizers, traditional
convex optimization theories and methods for solving  this nonconvex problem are very difficult. Actually,
  many nonconvex minimization problems in global optimization are considered to be NP-hard (see Gao and Sherali, 2009).

 Canonical duality theory is a  potentially useful methodological concept which was developed originally from complementary
 variational problems in  nonconvex mechanics (see Gao and Strang, 1989). This theory is
 composed mainly of (1) a canonical dual transformation method, which can be used to reformulate the nonconvex
  primal problem as a perfect dual problem without duality gap;
  (2) a complementary-dual principle, which provides an analytical solution form to the primal problem;
    (3) a triality theory, which can be used to identify both global and local extrema.
    This theory has been used successfully for solving a large class  of nonconvex/nonsmooth/discrete  problems in
   computational biology, global optimization, phase transitions of solids,
    finite element methods for post-buckling of large deformed structures, and Euclidian distance geometry problems
(see Gao et al, 2000 - 2012). The goal of this paper is to apply this theory for solving the nonconvex
minimization problem $(\mathcal{P}_2 )$.
In the next section, we  first introduce briefly the canonical duality theory with  a simple example of a one dimensional
double-well function optimization problem.
Then we use the canonical dual transformation to construct the canonical dual problem in
Section 3. Furthermore, the form of the analytical solution is
obtained from the criticality condition in Section 4.
In Section 5,
we present some numerical experiments.
Concluding remarks are given in the last section.

\section{Canonical duality theory: A brief review}
The basic idea of the canonical duality theory can be   demonstrated
by solving the following
general nonconvex problem (the primal problem $(\calP)$ in short)
\begin{eqnarray}
(\calP): \; \min_{ \bx \in \calX_a}
\left\{ \PP(\bx) = \half   \la \bx ,  \bQ \bx  \ra  -
\la  \bx ,  \bff \ra  +  \WW( \bx) \right\},
\end{eqnarray}
where
$ \bQ  \in \real^{n\times n} $ is  a given symmetric indefinite matrix,
$\bff \in \real^n$ is a given vector,  $\la \bx, \bx^* \ra $
denotes the bilinear form
between $\xx$ and its dual variable $\bx^*$,
$\WW(\bx)$ is a general nonconvex function; and $\calX_a \subset \real^n$ is a
  given  feasible space.

The  \textbf{key step} in  the canonical dual  transformation
is to choose a nonlinear operator,
\begin{eqnarray}
\bveps = \Lam (\bx):\calX_a \rightarrow \calE_a \subset \real^p
\end{eqnarray}
and a {\em canonical function} $\VV: \calE_a \rightarrow \real$
such that the nonconvex functional $\WW( \bx)$
can be recast by adopting a canonical form
$\WW( \bx) = \VV(\Lam(\bx))$.
Thus, the primal problem $(\calP)$  can be
written in the following canonical form:
\begin{eqnarray}
(\calP): \;  \min_{\bx \in \calX_a}
\left\{ \PP(\bx) =   \VV(\Lam(\bx)) - \UU(\bx)\right\},
\label{eq-canform}
\end{eqnarray}
where $\UU(\bx) =  \la \bx, \bff \ra - \half \la \bx , \bQ \bx \ra$.
By the definition introduced in
\citeasnoun{gao-book00}, a differentiable function $\VV(\bveps)$ is said to be
a \textit{canonical function}  on its domain  $\calE_a$ if the
duality mapping $\bvsig = \nabla \VV(\bveps)$ from $\calE_a$ to its range
$ \calS_a \subset \real^p $
is invertible. Let   $\la \bveps ;  \bvsig \ra $
denote  the bilinear form on $\real^p$.
Thus, for the given canonical function $\VV(\bveps)$,
its Legendre conjugate
$\VV^*(\bvsig)$ can be defined uniquely by the Legendre transformation
\begin{eqnarray}
\VV^*(\bvsig)  = \sta \{ \la \bveps ;  \bvsig \ra - \VV(\bveps ) \; | \; \; \bveps \in \calE_a \},
\end{eqnarray}
where  the notation $\sta \{ g(\bveps) | \; \bveps \in \calE_a\}$
stands for finding stationary point of $g(\bveps)$ on $\calE_a$.
It is easy to prove that
the following canonical  duality relations hold on $ \calE_a \times \calS_a$:

\begin{eqnarray}
\bvsig =\nabla \VV(\bveps) \; \Leftrightarrow \;  \bveps = \nabla
\VV^*(\bvsig) \; \Leftrightarrow
\VV(\bveps) + \VV^*(\bvsig) =
\la \bveps ; \bvsig \ra  . \label{eq-candual}
\end{eqnarray}
By this one-to-one canonical duality,   the nonconvex term
$W( \bx)=\VV(\Lam(\bx))$ in the problem $(\calP)$ can be replaced by
$\la \Lam(\bx) ; \bvsig \ra -
\VV^*(\bvsig)$ such that the nonconvex
function $\PP(\bx)$ is reformulated  as
the so-called Gao and Strang total complementary function:
\begin{eqnarray}
\Xi(\bx, \bvsig) = \la  \Lam(\bx) ; \bvsig \ra    - \VV^*(\bvsig) -  \UU(\bx).
\label{eq:xi}
\end{eqnarray}
By using this total complementary function,
the canonical dual function $\PP^d (\bvsig)$
can be obtained   as
\begin{eqnarray}
\PP^d(\bvsig) &=& \sta \{ \Xi(\bx, \bvsig) \; | \; \bx \in \calX_a   \} \nonumber \\
&=& \UU^\Lam(\bvsig) - \VV^*(\bvsig),
\end{eqnarray}
where $\UU^\Lam(\bx)$ is defined by
\begin{eqnarray}
\UU^\Lam(\bvsig) =  \sta \{ \la  \Lam(\bx) ;  \bvsig \ra - \UU(\bx) \;
| \;\; \bx \in \calX_a \}.
\end{eqnarray}
In many applications, the geometrically nonlinear operator
$\Lam(\bx)$ is usually  quadratic function
\begin{eqnarray}
\Lam(\bx)=\half \la \bx, D_k \bx \ra +\la \bx, \bb_k \ra,
\end{eqnarray}
where $D_k \in \real^{n \times n}$ and
$\bb_k\in \real^n  (k = 1, \cdots, p)$ are given. Let $\bvsig= [\vsig_1,\cdots, \vsig_p]^T$.
In this case, the canonical dual function can be written in the following form:
\begin{eqnarray}
\PP^d(\bvsig)=-\half \la \bF(\bvsig), \bG^{-1}(\bvsig) \bF(\bvsig) \ra -V^{\ast}(\bvsig),
\end{eqnarray}
where $\bG(\bvsig) = \bQ +\sum_{k=1}^p \bvsig_k D_k$, and
$\bF(\bvsig)=\bff-\sum_{k=1}^p \vsig_k \bb_k$.

Let
\[
\calS^+_a = \{\bvsig\in \real^p|\; G(\bvsig) \succ 0\}.
\]
Therefore, the canonical dual problem is proposed as
\begin{eqnarray}
(\calP^d): \;\;   \max \{ \PP^d(\bvsig) | \;\; \bvsig \in \calS^+_a\} \vspace{-.3cm} ,
\end{eqnarray}
which is a concave maximization problem over a convex set
$\calS^+_a \subset \real^p$.

\begin{thm}[\citeasnoun{gao-book00}]\label{duality}
Problem $(\calP^d)$ is canonically dual to $(\calP)$ in the sense that
if $\barbvsig$ is a critical point of $\PP^d(\bvsig)$, then
\begin{eqnarray}\label{eq-anasol}
\barbx = \bG^{-1}(\barbvsig) \bF(\barbvsig)
\end{eqnarray}
is a critical point of $\PP(\bx)$ and
\begin{eqnarray}
\PP(\barbx) =  \Xi(\barbx, \barbvsig) = \PP^d(\barbvsig). \label{eq-p=d}
\end{eqnarray}
If $\barbvsig \in \calS^+_a $ is a solution to $(\calP^d)$, then $\barbx  $
is a global minimizer of $(\calP) $ and
\begin{eqnarray}
\min_{\bx \in \calX_a } \PP(\bx) =  \Xi(\barbx, \barbvsig)=
\max_{\bvsig \in \calS_a^+} \PP^d(\bvsig)\label{trisaddle}.
\end{eqnarray}
Conversely, if $\barbx$ is a solution to $(\calP)$, it must be in the form of
(\ref{eq-anasol}) for critical solution $\barbvsig$ of $\PP^d(\bvsig)$.
\end{thm}

\begin{figure}[!t]
\centering
\includegraphics[width=2.5in]{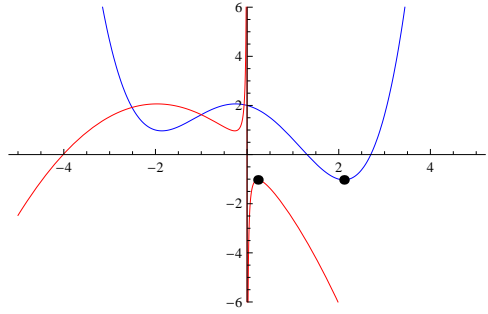}
\caption{Graphs of the primal function $\PP(\bx)$  (blue)
and its canonical dual function $\PP^d(\vsig)$ (red).}
\label{onedim}
\end{figure}

To help explain the theory, we   consider a simple nonconvex optimization in
$\real^n$:
\begin{eqnarray}
\min \PP(\bx)=\half \alpha (\half\|\bx\|^2-\lam)^2-\bx^T \bff, \; \forall \bx \in \real^n,
\end{eqnarray}
where $\alp, \lam > 0$ are given parameters.
The criticality condition $\nabla P(\bx)=0$ leads to a nonlinear algebraic
equation system in $\real^n$
\begin{eqnarray}
\alpha (\half \|\bx\|^2-\lam)\bx =\bff.
\end{eqnarray}
Clearly, it is difficult  to solve this nonlinear algebraic equation directly.
Also traditional convex optimization theory
 can't be used to identify global minimizer.  However, by the
canonical dual transformation, this problem can be solved.
To do so, we let  $\bxi=\Lam(u)=\half\|\bx\|^2-\lam \in \real$. Then,
the nonconvex function $W(\bx) = \half \alpha(\half \| \bx \|^2 -\lam)^2$
can be written in canonical form
$V(\bxi) = \half \alpha \bxi^2$.
Its Legendre conjugate is  given by
$V^{\ast}(\vsig)=\half \alpha^{-1}\vsig^2$, which is strictly convex.
Thus,
the total  complementary function for this nonconvex optimization
problem is
\begin{eqnarray}
\Xi(\bx,\vsig)=(\half \|\bx\|^2 - \lam) \vsig-\half
\alpha^{-1}\vsig^2 - \bx^T \bff.
\end{eqnarray}
For a fixed $\vsig \in \real$, the  criticality condition
$\nabla_{\bx} \Xi(\bx)=0$ leads to
\begin{eqnarray}\label{balance}
\vsig \bx-\bff=0.
\end{eqnarray}
For each  $\vsig \neq 0  $,
the  equation
(\ref{balance}) gives $\bx=\bff/\vsig$ in vector form. Substituting this into the
total complementary function $\Xi$,
the canonical dual function can be easily obtained as
\begin{eqnarray}
\PP^d(\vsig)&=&\{\Xi(\bx,\vsig)| \nabla_{\bx} \Xi(\bx,\vsig)
=0\}\nonumber\\
&=& -\frac{\bff^T \bff}{2 \vsig}-\half \alpha^{-1} \vsig^2
-\lam \vsig, \;\;\; \forall \vsig\neq 0,
\end{eqnarray}
which has only one variable!
The critical point of this canonical function is obtained
by solving the following dual algebraic  equation
\begin{eqnarray}
(\alpha^{-1} \vsig+\lam)\vsig^2=\half \bff^T \bff.
\end{eqnarray}
For any given parameters $\alpha$, $\lam$ and the vector $\bff\in \real^n$,
this cubic algebraic equation can be solved analytically to have  at most three real roots
satisfying $\vsig_1 \ge 0 \ge \vsig_2\ge \vsig_3$,
and each of these roots leads to a critical point of the nonconvex function
$P(\bx)$, i.e., $\bx_i=\bff/\vsig_i$, $i=1,2,3$. By
the fact that $\vsig_ 1 \in \calS^+_a = \{ \vsig \in \real\; |\; \vsig > 0 \}$,
then Theorem 1 tells us that $\bx_1$ is
a global minimizer of $\PP(\bx)$.
Consider one dimension problem with $\alpha= 1$, $\lam=2$, $f= \half$,
the primal function and canonical dual function
are shown in Fig. \ref{onedim}, where,  $x_1= 2.11491$ is global minimizer
of $P(\bx)$,
$\vsig_1=0.236417$ is global maximizer of $\PP^d(\bvsig)$, and
$\PP(x_1)=-1.02951=\PP^d(\vsig_1)$ (See the two black dots).

The canonical duality theory was original developed for handling general
nonconvex systems (see \citeasnoun{gao-book00}). The canonical dual transformation can be used to convert
a nonconvex problem into a canonical dual problem without duality gap, while the
classical dual approaches may suffer from having a potential gap
(see \citeasnoun{rock74}). The complementary-dual principle provides
a unified form of analytical solutions (\ref{eq-anasol}) to general
nonconvex problems in  continuous or discrete systems.

\section{Canonical Dual Approach for Solving Nonconvex Problem $(\mathcal{P}_2 )$}
In order to solve the nonconvex minimization problem $(\mathcal{P}_2 )$ by the canonical duality theory,
 we  need to fix the parameters $q, r, K$ first and introduce a  geometrical nonlinear measure
 $\Lam_1 : \real^n \rightarrow \real^{n-1}$ defined by
\begin{eqnarray*}
\bxi=\Lam_1(\by)=r \by^T \bA \by+(D-(1+r)R)\by+\frac{C}{K}  .
\end{eqnarray*}
The canonical function associated with this nonlinear measure  is a
quadratic function: $V_1(\bxi)=\half \alpha \|\bxi\|^2 $  and the duality relation
\begin{eqnarray*}
\bvsig=\nabla V_1(\bxi)=\alpha \bxi
\end{eqnarray*}
is invertible.
Then the Legendre conjugate is defined by
\begin{eqnarray*}
V_1^{\ast} =\max\{ \bxi^T \bvsig-V_1(\bxi) |   \bxi \in \real^{n-1} \}=\frac{1}{2 \alpha}\bvsig^T \bvsig.
\end{eqnarray*}

On the same time, we rewrite the linear inequality constraints $0< y_i< 1$,
$i=1,\cdots,n$  in the canonical form:
\[
\by \circ (\by -\be) <  0 ,
\]
where $\be = ( 1, \dots, 1) \in \real^n$ is an one-vector,
the notation $\bs \circ \bt := (s_1 t_1, s_2 t_2, \cdots, s_n t_n)$ denotes
the Hadamard product for any two vectors $\bs,\bt \in \real^n$.
Introduce a quadratic geometrical operator $\Lam_2:\real^n \rightarrow \real^n$
\begin{eqnarray*}
\beps=\Lam_2(\bx)=\bx\circ(\bx-\be),
\end{eqnarray*}
and indicator function
\begin{eqnarray*}
V_2(\beps)= \left\{ \begin{array}{ll}
0 & \mbox{ if } \beps <  0,   \\
+ \infty & \mbox{ otherwise},
\end{array} \right.
\end{eqnarray*}
then  we know that, by theory of convex analysis, the subdifferential $\bsig \in \partial V_2(\beps)$
is equivalent to the following KKT conditions:
\[
\beps < 0 , \;\; \bsig \ge 0, \;\; \beps^T \bsig = 0.
\]
Therefore, for the fixed parameter $q, r, K$,  the inequality constrained primal problem $(\mathcal{P}_2 )$ can be written in the following canonical form.
\begin{eqnarray*}
 (\mathcal{P}_3 )~~~~ \min~~ \left\{ \PP_3 (\by)= V_1(\Lam_1(\by)) + V_2(\Lam_2(\by)) +
\left \|q\by-\frac{\II}{K} \right \|^2 \; | \;  \by \in \real^n \right\} .
\end{eqnarray*}

 Since the indicator function $V_2$ is nonsmooth, its conjugate is defined by the  Fenchel transformation
\begin{eqnarray*}
V^{\sharp}_2 (\bsig)=\sup \{
\beps^T \bsig-V_2(\beps) \; | \beps \in \real^n  \} = \left\{ \begin{array}{ll}
0 \; & \mbox{ if } \bsig \ge 0 \\
+ \infty & \mbox{otherwise}.
\end{array} \right.
\end{eqnarray*}
Thus, the primal function can be reformulated as the total
complementary function:
\begin{eqnarray*}
\Xi(\bx,\bvsig,\bsig)&=&\bvsig^T\Lam_1(\bx)- V_1^{\ast}(\bvsig)+
\bsig^T\Lam_2(\bx)- V_2^{\sharp}(\bsig)+\|q \bx-\frac{I}{K} \|^2\\
&=& \half \bx^T(\bvsig, \bsig) \bG(\bvsig,\bsig) \bx(\bvsig, \bsig)-
\bF^T(\bvsig, \bsig)\bx-V_1^{\ast}(\bvsig)-V_2^{\sharp}(\bsig)
\end{eqnarray*}
in which
\begin{eqnarray*}
\bG(\bvsig, \bsig) &=& 2( r \bA^T  \bvsig + q^2 \HH + \Diag(\bsig)),\\
\bF(\bvsig, \bsig) &=& 2 \frac{q}{K} \II+ \bsig - (\DD -(1+\rr) \RR)^T
\bvsig,
\end{eqnarray*}
where $\HH \in \real^{n\times n} $ is an identity matrix, and
$\Diag(\bsig) \in \real^{n \times n}$ denotes a diagonal matrix with
$\{\sig_i\} \; (i=1, \cdots, n)$ as its diagonal entries.
For a fixed $\bvsig, \bsig$, the criticality condition
$\nabla_{\bx}\Xi(\bx,\bvsig,\bsig)=0$ lead to the following
canonical equilibrium equation:
\begin{eqnarray*}
\bG(\bvsig,\bsig) \bx =\bF(\bvsig,\bsig),
\end{eqnarray*}
This equation can be solved analytically  in the form of
\begin{eqnarray}
\bx=\bG^{-1}(\bvsig,\bsig) \bF(\bvsig,\bsig) \label{xformu}
\end{eqnarray}
on the canonical dual feasible space defined by
\begin{eqnarray*}
\calS_a= \{ (\bvsig, \bsig) \in \real^{(n-1) + n}|\;\bsig \geq 0, \;\;
\det G(\bvsig,\bsig) \neq 0 \}.
\end{eqnarray*}
Thus,  the canonical dual problem can be formulated as
\begin{eqnarray*}
(\mathcal{P}^d_3):  \;\sta\left\{P^d_3(\bvsig,\bsig)= -\half \bF(\bvsig,
\bsig)^T \bG^{-1}(\bvsig,\bsig) \bF(\bvsig, \bsig)-\frac{1}{2
\alp}\bvsig^T \bvsig + \frac{1}{K}\CC^T \bvsig+ \frac{1}{K^2}\II^T
\II  \; | \; \;\bvsig \in \calS_a\right\}.
\end{eqnarray*}

\begin{thm}[Complementary-Dual Principle] \label{duality}
For the given parameters  $q$, $K$, and  $r$, if $(\barbsig, \barbvsig) $ is a KKT point of
$(\mathcal{P}^d_3)$, then the vector
\begin{eqnarray}
\barby= \bG^{-1}(\barbvsig,\barbsig)\bF(\barbvsig,\barbsig) \label{solution3}
\end{eqnarray}
is a KKT
point of $(\mathcal{P}_3 )$, and $P_3 (\barby)=P^d_3(\barbvsig,\barbsig  )$.
\end{thm}
{\em Proof}. Suppose that $(\barbvsig, \barbsig)$ is a KKT point of
$(\mathcal{P}^d_3)$, by introducing Lagrange multiplier $\epsilon$ to relax
the inequality condition $\barbsig \ge 0 $ of the dual problem
$(\mathcal{P}^d_3)$,   we have
\begin{eqnarray}
\nabla_{\bvsig} P^d_3(\barbvsig,\barbsig) =
r\barby(\barbvsig,\barbsig)^T \bA \barby(\barbvsig,\barbsig) + B
\barby(\barbvsig,\barbsig)
+ \frac{C}{K}- \frac{1}{\alp} \barbvsig &=& 0,\label{partione}\\
\nabla_{\bsig} P^d_3(\barbvsig,\barbsig) = \barby(\barbvsig,\barbsig)
\circ \barby(\barbvsig,\barbsig)
-\barby(\barbvsig,\barbsig) &=& \beps,\label{partitwo}\\
\barbsig \geq 0,\; \;\; \beps &<& 0, \label{parti3}\\
\barbsig^T \beps &=& 0\label{parti4}.
\end{eqnarray}
Therefore, the equality (\ref{partione}) gives $\barbvsig=\alp(r \barby^T \bA \barby + B \barby +
\frac{C}{K})$, where  $B= D-(1+ r)R$.

By the equation   (\ref{partitwo}) and the KKT conditions (\ref{parti3}) and the complementarity condition
(\ref{parti4}), we have
\begin{eqnarray}
\nabla P_3 (\barby)+2 \Diag(\barbsig) \barby - \barbsig &=& 0\\
0<\barby <1,\; \barbsig &\geq & 0\\
\barbsig(\barby \circ \barby -\barby) &=& 0,
\end{eqnarray}
where
\begin{eqnarray*}
\nabla P_3 (\barby)= (2 r \bA^T \barbvsig +2 q^2 H) \barbx -(2 q
\frac{\II}{K}-\BB^T \barbvsig).
\end{eqnarray*}
Therefore,
\begin{eqnarray*}
\barby &=& (2(r \bA^T \barbvsig + q^2 H + \Diag( \barbsig))) ^{-1}
(2q \frac{\II}{K }+ \barbsig -B^T \barbvsig)
\end{eqnarray*}
is a KKT point of the primal problem $(\mathcal{P}_3 )$.

Moreover, in terms of $\barby=\bG^{-1} (\barbvsig, \barbsig)
\bF(\barbvsig, \barbsig)$, we have
\begin{eqnarray*}
&&P^d_3(\barbvsig,\barbsig) \\
&=& - \half \bF(\barbvsig,\barbsig)^T \bG^{-1}(\barbvsig, \barbsig)
\bF(\barbvsig,\barbsig) -\frac{1}{2 \alp}
\barbvsig^T \barbvsig +\frac{1}{K} C^T \barbvsig +\frac{1}{K^2} \II^T \II\\
&=& \half \bx^T (2( r \bA^T \barbvsig + q^2 H + \Diag(\barbsig)))
\barbx- \barbx^T (2 q \frac{\II}{K}+ \barbsig -B^T \barbvsig)
-\frac{1}{2 \alp}
\barbvsig^T \barbvsig +\frac{1}{K} C^T  \barbvsig +\frac{1}{K^2} \II^T \II\\
&=& \barbvsig^T (r \barby^T \bA \barby) +( B\barby)^T \barbvsig +
\frac{1}{K} C^T \barbvsig -\frac{1}{2 \alp} \barbvsig^T \barbvsig +
\barby^T (q^2 H) \barby -2 (q\barby)^T
\frac{\II}{K}+ \frac{1}{K^2} \II^T \II \\
&=& \frac{\alp}{2} \| r \barby ^T \bA \barby + B \barby+
\frac{C}{K}\|^2 +\|q \barby -\frac{\II }{K}\|^2 = P_3 (\barby)
\end{eqnarray*}
This proves the theorem. \hfill $\Box$

For further discussion on extremality properties of the
solution (\ref{solution3}),
we introduce the following feasible space
\begin{eqnarray}
&&\calS_a^+=\{(\bvsig, \bsig) \in \calS_a \; | \;\; \bG (\bvsig, \bsig) \succ 0\}.
\end{eqnarray}
Then, the extremality of the solution (\ref{solution3}) can be identified by the following theorem.
\begin{thm}\label{opcr}
For the given parameters $q, K, r$, suppose that $(\barbvsig, \barbsig) \in \calS_a^+$ is a critical point of the canonical dual function
$\PP^d_3(\barbvsig, \barbsig)$ and $\barby = \bG^{-1}(\barbvsig, \barbsig) \bF(\barbvsig,\barbsig) $. Then,
$\barby$ is a global minimizer of $\PP_3(\by) $
on $\real^{n}$ if and only if
$(\barbvsig, \barbsig)$ is a global maximizer of
$\PP^d_3(\barbvsig, \barbsig)$ on $\calS_a^+$, i.e.,
\begin{eqnarray}
\PP_3(\barby)=\min_{\by\in {\real^{n}}} \PP_3(\by)
\Leftrightarrow\max_{(\bvsig,\bsig)\in \calS_a^+}
\PP^d_3(\bvsig,\bsig)=\PP^d_3(\barbvsig,\barbsig).
\end{eqnarray}
\end{thm}

\begin{remark}
Whence the global optimal solution
  $\barby$ of the nonconvex  problem $(\mathcal{P}_3 )$ is obtained by the
  canonical dual approach,  the optimal parameters $\bar{q}, \bar{r}, \bar{K}$ can be obtained by solving the
  following minimization problem
  \begin{eqnarray*}
  &(\mathcal{P}_4 )~~~~&\min~~\PP_4( q, K,
r)= \left \|q\barby-\frac{\II}{K} \right \|^2+\frac{\alp}{2}
\left \|r \barby^T \bA \barby +(\DD-(1+ r)\RR)\barby+ \frac{\CC}{K} \right \|^2\\,
&& {\rm s.t.}~~ q>0, \; K>0, \; r>0 .
\end{eqnarray*}
 By combining   together problems $(\mathcal{P}_3 )$
  and $(\mathcal{P}_4 )$, the global optimal solution
  $(\barby, \bar{q},
  \bar{r}, \bar{K})$ of the problem $(\calP_2)$ can be obtained by certain alternative  iteration method.
  Then, the  global minimizer to the problem  $(\mathcal{P}_1)$ is $\barbx= \bar{K} \barby$ and
we have  $P_1(\barbx, \bar{q}, \bar{K}, \bar{r})= \bar{K}^2 P_2 (\barby, \bar{q}, \bar{K}, \bar{r})$.
\end{remark}

\section{Applications}

We now list a few examples to illustrate the applications of the
theory presented in this paper.

{\bf Example 1}.  The data we used in Table 1 is  from Quinn T and Deriso
R.B.'s book (see \citeasnoun{qude99}).

\begin{table}[!b] 
\tblcaption{Yield (Catch) and
catch-per-unit-effort ${\bf I}$ (CPUE)  for each of year}
{\mbox{\tabcolsep=18pt\begin{tabular}{@{}ccc@{}}
\tblhead{ Year & $\bC$ (Catch) & ${\bf I}$ (CPUE)\\[-9.5pt]}\\[-9.5pt]\\
1&3706&82\\
2&2662&59\\
3& 2055&46\\
4& 1658&37\\
5& 1379&31\\
6& 1171&26\\
7& 1011&22\\
8& 884&20\\
9& 781&17\\
10& 696&15\\
11& 85&17\\
12& 111&22\\
13& 142&28\\
14& 178&36\\
15&215&43\\
16& 253&51\\
17& 289&58\\
18& 321&64\\
19& 349&70\\
20& 371&74\\
21& 1451&73\\
22&1341&67\\
23&1263&63\\
24& 1206&60\\
25&1162&58\\
26& 1129&56\\
27& 1103& 55 \lastline
\end{tabular}
}}
\end{table}

By the canonical duality  method presented in this paper,  the optimal parameters  we obtained are
$\barq=0.0112$,  $\barK=14777.4$,
$\barr=0.1875$, and the optimal solution $\barby$ to model
$(\mathcal{P}_2)$ is
\begin{eqnarray*}
\barby&=&\{0.87804,0.595806,0.465162,0.407183,0.368356,0.353525,0.353367,0.366521,0.368154,\\
      &&0.372081,0.360713,0.391465,0.409868,0.432046,0.445403,0.458043,0.467348,0.477866,\\
      &&0.490154,0.509345,0.542681,0.476886,0.43092,0.386888,0.351872,0.296357,0.245213\}.
\end{eqnarray*}
The optimal objective function value is therefore $P_2 (\barby, \bar{q}, \bar{K},\bar{r})=0.0698$.
Figure 2 presents a clear view of the tendency during these 27
years.
\begin{figure}[!t] 
\centering\includegraphics[width=0.45\textwidth] {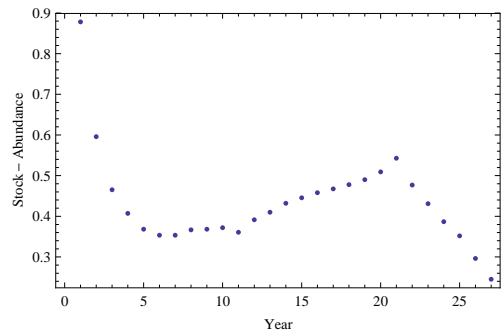}
\caption{Stock abundance during 27 years.}
\label{series1}
\end{figure}

{\bf Example 2} The data we used in this example  is  the New Zealand rock
lobster during   46 years  given in \citeasnoun{php93}.
\begin{table}[!b] 
\tblcaption{Catch and catch-rate data}
{\mbox{\tabcolsep=18pt\begin{tabular}{@{}cccccc@{}}
\tblhead{Year & Catch& CPUE&Year &Catch& CPUE\\[-9.5pt]}\\[-9.5pt]
1945& 809&3.49&1968&4975&1.53\\
1946& 854&3.38&1969&4786&1.32\\
1947& 919&3.18&1970&4699&1.45\\
1948& 1360&3.56&1971&4478&1.40\\
1949& 1872&1.79&1972&3495&1.09\\
1950& 2672&4.35&1973&3784&1.23\\
1951& 2834&2.33&1974&3643&1.12\\
1952& 3324&2.57&1975&2987&0.92\\
1953& 4160&2.88&1976&3311&1.02\\
1954& 5541&3.85&1977&3237&1.0\\
1955& 5909&4.16&1978&3418&1.05\\
1956& 6547&4.34&1979&4050&1.09\\
1957& 5049&3.70&1980&4190&1.02\\
1958& 4447&2.37&1981&4058&1.01\\
1959& 4018&2.46&1982&4331&0.98\\
1960& 3762&2.06&1983&4385&1.01\\
1961& 4042&2.21&1984&4911&0.85\\
1962& 4583&2.19&1985&4856&0.84\\
1963& 4554&2.44&1986&4657&0.81\\
1964& 4597&2.14&1987&4500&0.84\\
1965& 4984&2.18&1988&3128&0.68\\
1966& 5295&2.13&1989&3318&0.62\\
1967& 4782&1.86&1990&2770&0.54\lastline
\end{tabular}
}}
\end{table}

\begin{figure}[!t] 
\centering\includegraphics[width=0.45\textwidth] {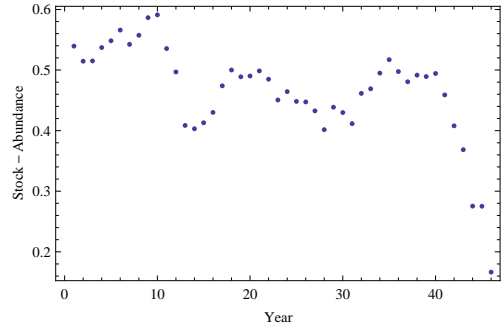}
\caption{Stock abundance during 46 years.}
\label{series2}
\end{figure}

By the canonical duality  method, the optimal parameters for the
given data are  $\barq=0.0005$, $\barK=28238.1$, $\barr=0.5729$, and the
optimal solution $\barby$ to model $(\mathcal{P}_2 )$ is
\begin{eqnarray*}
\barby&=&\{0.539323,0.514615,0.515001,0.537057,0.548343,0.566006,0.542414,0.557294,\\
      &&0.586471,0.591127,0.535325,0.496853,0.40859,0.403013,0.413048,0.429969,\\
      &&0.473993,0.49995,0.488957,0.490067,0.498439,0.484909,0.450408,0.464355,\\
      &&0.448175,0.447358,0.432635,0.401602,0.438707,0.429822,0.411511,0.461489,\\
      &&0.468958,0.494921,0.517036,0.497474,0.480722,0.491527,0.489311,0.49443,\\
      &&0.459034,0.40785,0.368474,0.275356,0.275089,0.166475\}.
\end{eqnarray*}
The optimal objective function value  $P_2(\barby, \bar{q}, \bar{K}, \bar{r})=2.1596$
(see Figure \ref{series2}).

\section{Concluding remarks}
We have presented a canonical dual approach  for solving a challenging population growth problem
in discrete dynamical systems. By using the finite difference and the least squares methods,
this nonlinear problem is reformulated as a nonconvex  optimization problem with inequality constraints.
The problem has extensive applications in   ecology, neural networks, medicine, economics,  statistical physics, and
much more. Due to the nonconvexity of the target function with unknown parameters,
the problem is fundamentally difficult and can't be solved  by
traditional direct methods.
However, by the canonical duality theory, global optimal solution is obtained for the first time
to this well-known  problem over the whole time domain.
 Applications illustrated that this theory is efficient for solving
 large-scale population growth problems.

\vspace{1.0 cm}

\noindent{\bf Acknowledgement}:
This paper was partially supported by a grant (AFOSR FA9550-10-1-0487)
from the US Air Force Office of Scientific Research. Dr. Ning Ruan was
supported by a funding from the Australian Government
under the Collaborative Research Networks (CRN) program.

\end{document}